\documentclass[fleqn,usenatbib]{mnras}

\usepackage{newtxtext,newtxmath}
\usepackage[authoryear]{natbib}
\usepackage{hyperref}
\usepackage[T1]{fontenc}

\DeclareRobustCommand{\VAN}[3]{#2}
\let\VANthebibliography\thebibliography
\def\thebibliography{\DeclareRobustCommand{\VAN}[3]{##3}\VANthebibliography}

\usepackage{graphicx}	% Including figure files
\usepackage{amsmath}	% Advanced maths commands
\newcommand{\beq}{\begin{equation}}
\newcommand{\eeq}{\end{equation}}

\title[Discrete Dynamical Friction]{A New Discrete Dynamical Friction Estimator Based on $N$-body Simulations}

\author[Ma et al.]{
Linhao Ma$^{1,}$\thanks{E-mail: lma3@caltech.edu},
Philip F. Hopkins$^{1}$,
Luke Zoltan Kelley$^{2}$
and Claude-André Faucher-Giguère$^{2}$
\\
$^{1}$TAPIR, Mailcode 350-17, California Institute of Technology, Pasadena, CA 91125, USA\\
$^{2}$CIERA and Department of Physics \& Astronomy, Northwestern University, Evanston, IL 60208, USA
}

\date{Accepted 2023 January 3. Received 2022 December 6; in original form 2022 August 24}

\pubyear{2020}

\begin{document}
\label{firstpage}
\pagerange{\pageref{firstpage}--\pageref{lastpage}}
\maketitle

% Abstract of the paper
\begin{abstract}
A longstanding problem in galactic simulations is to resolve the dynamical friction (DF) force acting on massive black hole particles when their masses are comparable to or less than the background simulation particles. Many sub-grid models based on the traditional Chandrasekhar DF formula have been proposed, yet they suffer from fundamental ambiguities in the definition of some terms in Chandrasekhar's formula when applied to real galaxies, as well as difficulty in evaluating continuous quantities from (spatially) discrete simulation data. In this work we present a new sub-grid dynamical friction estimator based on the discrete nature of $N$-body simulations, which also avoids the ambiguously-defined quantities in Chandrasekhar's formula. We test our estimator in the {\small GIZMO} code and find that it agrees well with high-resolution simulations where DF is fully captured, with negligible additional computational cost. We also compare it with a Chandrasekhar estimator and discuss its applications in real galactic simulations.
\end{abstract}

% Select between one and six entries from the list of approved keywords.
% Don't make up new ones.

%%%%%%%%%%%%%%%%%%%%%%%%%%%%%%%%%%%%%%%%%%%%%%%%%%

%%%%%%%%%%%%%%%%% BODY OF PAPER %%%%%%%%%%%%%%%%%%

\section{Introduction}
\label{sec:intro}

An essential element in the study of galactic dynamics is the process of dynamical friction (DF, \citealt{chandrasekhar1943}), a statistical effect of numerous two-body scatterings which causes a massive particle to lose its momentum when it travels through a medium of much lighter background particles. DF is believed to be an important effect to drive massive black holes (BHs, from intermediate mass BHs to super-massive BHs) to galactic centers (see, e.g. \citealt{Ostriker1999,Weller2022,Chen2022}), and it plays an essential role in the evolution of globular clusters (see, e.g. \citealt{PortegiesZwart2002,Gurkan2004,Alessandrini2014,Shi2021}). Hence the evaluation of DF is important in studying the evolution of galaxies, globular clusters, and black holes in a wide variety of contexts.

In numerical $N$-body simulations with sufficient resolution (such as in the limit in which all bodies such as stars, black holes, or even dark matter particles are represented by individual $N$-body particles), DF will be automatically captured. However, as DF is an accumulated effect of many weak encounters in the regime where the ``target'' mass is much larger than the mass of the ``background'' particles masses ($M_\mathrm{target\;particle}\gg M_\mathrm{background\;particle}$), it is often not possible to fully resolve this background. This is especially true in large-scale simulations of e.g.\ galactic scales, where a typical ``$N$-body particle'' can easily have mass much larger than intermediate and super-massive black holes ($\gg 10^{4}\,M_{\odot}$), let alone the masses of individual stars, dark matter particles, or hydrogen ions. Specifically, when the $N$-body particle mass becomes comparable to or larger than the ``target'' mass, the explicit results of an $N$-body solver will not return the correct DF forces. For example, in e.g.\ the ``high-resolution'' simulations of high-redshift galaxies in \citet{xiangchengma2018,xiangchengma2017,xiangchengma2019}, the baryonic mass resolution is $\Delta m_i\sim7000m_\odot$ and the dark matter mass resolution is 5 times larger, which makes it impossible to resolve dynamical friction effects for BHs or other ``sink'' particles (e.g.\ particles which might represent unresolved massive, dense structures such as globular clusters, or hyper-dense exotic dark matter structures, etc.) less massive than $\sim10^5 M_\odot$. Hence, in these types of simulations, an additional ``sub-grid'' DF force must be added to these ``target'' particles to attempt to recover their real dynamics, to replace the lost information of individual two-body encounters in the smoothed-out gravity potential in simulations.

Multiple sub-grid DF models have been proposed in the literature (e.g. \citealt{colpi:2007.binary.in.mgrs,dotti:bh.binary.inspiral,Tremmel2015,Pfister2019}) based on the classical Chandrasekhar's dynamical friction formula (\citealt{chandrasekhar1943}, or C43 hereafter):
\beq
\label{eqn:chandra_original}
{\bf a}_\mathrm{df}^\mathrm{C43}=-4\pi G^2Mm\ln{\Lambda}\int d^3{\bf v}_m f({\bf v}_m)\frac{{\bf v}_M-{\bf v}_m}{|{\bf v}_M-{\bf v}_m|^3}
\eeq
where $M$ and $m$ are the masses of the moving ``target'' particle and the background or field particles, respectively. Here ${\bf v}_M$ and ${\bf v}_m$ are their velocities, and $\Lambda$ is the Coulomb logarithm defined by $\Lambda\equiv b_\mathrm{max}/b_\mathrm{min}$ where $b_\mathrm{max}$ and $b_\mathrm{min}$ are the maximum and minimum impact factors of scattered particles in weak encounters. $f({\bf v}_m)$ is the velocity distribution of field particles, and, with the usual assumption of a Maxwellian velocity distribution with dispersion $\sigma$, the formula reduces to \citep{BinneyTremaine}:
\beq
\label{eqn:chandra_maxwell}
{\bf a}_\mathrm{df}^\mathrm{C43}=-\frac{4\pi^2G^2M\rho\ln{\Lambda}}{V_M^3}\bigg[\mathrm{erf}\bigg(\frac{v_m}{\sqrt{2}\sigma}\bigg)-\sqrt{\frac{2}{\pi\sigma}}e^{-v_m^2/2\sigma^2}v_m\bigg]{\bf v}_M
\eeq
i.e. the DF acceleration is proportional to the local field particle density $\rho$ and is in the opposite direction of the particle velocity ${\bf v}_M$, effectively acting as a ``friction'' force. Despite its elegance and (often surprising) accuracy in estimating the DF, Chandrasekhar's formula suffers from the following shortcomings when applied as a sub-grid model:

\begin{enumerate}

\item In deriving the formula, C43 assumes an isotropic and homogeneous medium of field particles. This is generally not true for real galaxies. For example, it has been pointed out that high-redshift galaxies and low-redshift dwarf galaxies could be chaotic and clumpy (e.g., \citealt{Weisz2014,meng2020,Velazquez2021}). The existence of such systems makes the physical assumptions behind C43 formula questionable.

\item The Coulomb logarithm is ambiguously defined, and is often selected ad-hoc in practice, with a case-dependent selection of the minimum and maximum impact parameters (see, e.g., \citealt{Tremmel2015,Pfister2019}), which introduces a large systematic uncertainty in the sub-grid model.

\item The formula has an explicit dependence on the local mass density, which must be evaluated from discrete $N$-body data for collisionless fluid (stars or dark matter, often ``blended'' with gas for which the density is continuously defined, depending on the numerical hydrodynamic method). The choice of how to do so is arbitrary and has no defined ``preferred'' scale. Most commonly it is done with a local kernel density estimator at some multiple of the resolution scale (see, e.g. \citealt{Tremmel2015}), but this is known to be quite noisy, and is not consistent with the unique local gas density available from hydrodynamic calculations.

\item The velocity integral and $f({\bf v}_m)$ must be estimated with some similar ad-hoc local estimator, which is also undefined, and different choices can lead to different {\em directions} for the dynamical friction acceleration. Usually the choice of a local kernel sampling amplifies numerical noise further here and means that $f({\bf v}_{m})$ must be assumed to be Maxwellian (since it cannot be fit to an arbitrary function given just a few local points). 

\item There is no self-consistent way to incorporate force softening, which is necessary in $N$-body simulations to avoid spurious divergences in the forces, as an $N$-body particle does not physically represent a point-mass particle. Fail to incorporate softening can produce inconsistent results between the (often softened) gravitational acceleration and the additional dynamical friction acceleration.

\item As C43 depends on {\em local} continuous field parameters but represents long-range forces, there is no way to self-consistently implement it in a way that conserves momentum, while in reality dynamical friction should be exactly conservative since it is derived from an infinite superposition of pair-wise $N$-body encounters;

\item Evaluating C43 numerically requires operations which are not algorithmically identical to the gravity solver in $N$-body equations, which introduces not only additional inconsistencies, but also substantial computational expense. This also means numerical convergence for C43 applied to $N$-body particles is undefined: there is no formal guarantee of convergence even on idealized, smooth problems.

\end{enumerate}

To tackle these problems, we develop a new sub-grid DF estimator which can be efficiently embedded into discrete $N$-body calculations in this work. The new estimator is based on a discrete version of the DF formula which can be applied to an arbitrary phase-space distribution of field particles, and avoids the fundamental ambiguity in the definitions of some terms in Chandrasekhar's formula. It also naturally embeds force softening and momentum conservation. It can also easily be generalized to assumptions beyond those of C43 for the nature of DF-like forces. We test our estimator in both on-the-fly simulations and in post processing, and compare our results to those from a Chandrasekhar DF estimator. The paper is written as follows: in \S~\ref{sec:formula} we derived our discrete DF formula. In \S~\ref{sec:methods} we describe the methods we use to test the estimator. In \S~\ref{sec:results} and \ref{sec:discussion} we present and discuss the results.

\section{Derivation of Our DF Formula}
\label{sec:formula}

Here we present the derivation of our discrete DF formula, and general comments on its application in $N$-body methods.

\begin{figure*}
    \centering
    \includegraphics[width=0.8\textwidth]{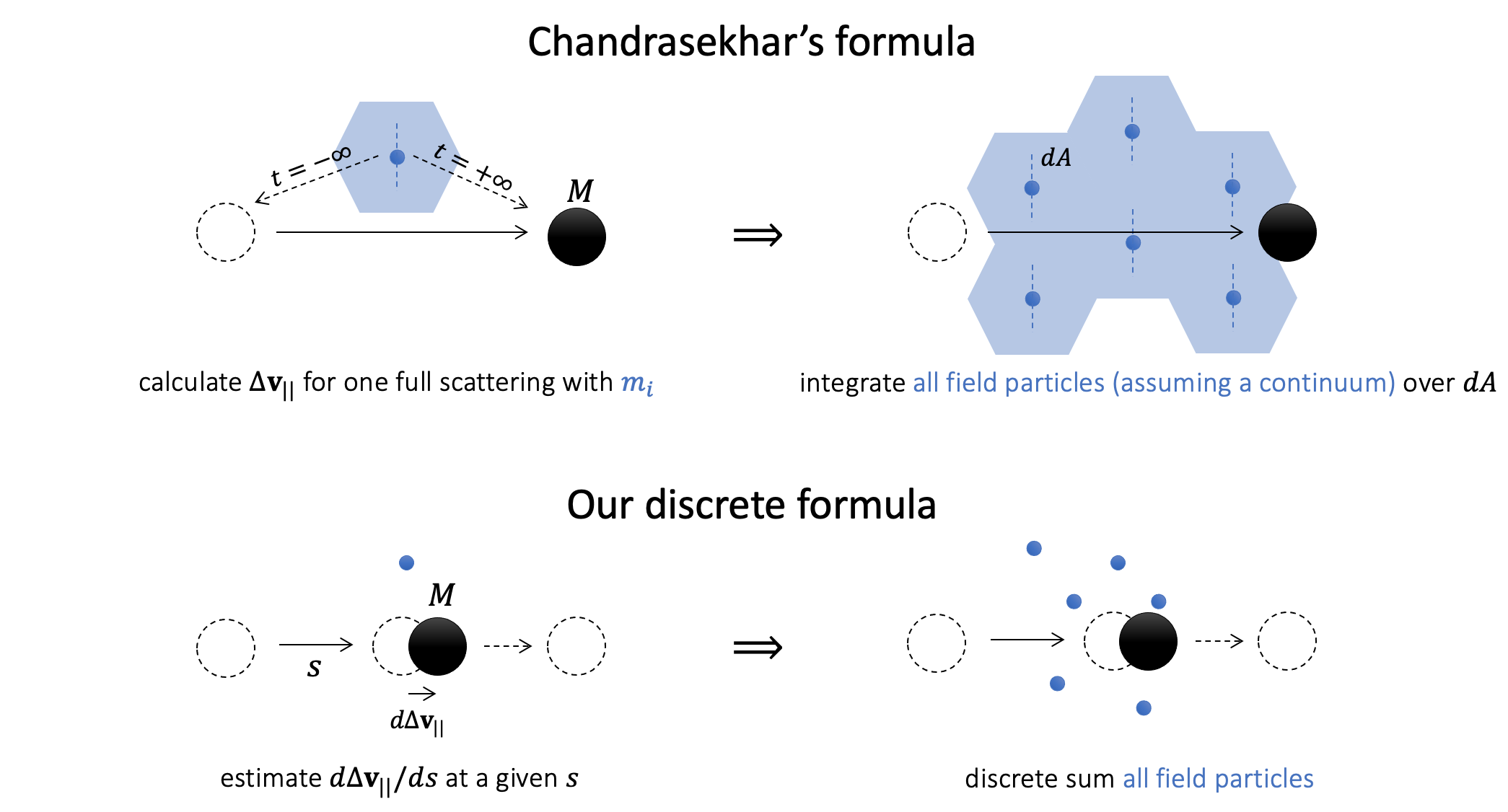}
    \caption{A comparison between the derivation of Chandrasekhar's DF formula (C43) and ours: in C43, Chandrasekhar calculates the change of velocity $\Delta {\bf v}_{||}$ for one full scattering, and integrates over the remaining two dimensions $dA$ (perpendicular to the direction of motion) for all field particles, assuming a homogeneous and isotropic continuum such that the overall contribution is characterized by the Coulomb logarithm; in our derivation, we estimate the change of velocity $d{\bf v}_{||}/ds$ over the line of motion (coordinated by $s$) at a given point in the scattering process, which allows us to integrate all field particles over the full configuration space (as the dimension along the line of motion is now recovered), such that a discrete numerical sum is possible.}
    \label{fig:derivation}
\end{figure*}

\subsection{Derivation}
\label{sec:derivation}

In C43, the classical DF formula is derived as follows: assume a test particle with mass $M$ travels through an infinite, homogeneous and isotropic medium (filled with background particles with mass $m\ll M$), and experiences a number of individual two-body encounters. During each encounter, along the direction of relative motion, the test particle velocity in the parallel direction to the initial relative velocity is altered by (after integrating along the encounter path $ds$ from $s\rightarrow-\infty$ to $s\rightarrow+\infty$)
\beq
\label{eqn:one_encounter}
\begin{split}
\Delta {\bf v}_{\|} &= \frac{2\,m\,{\bf V}}{M+m}\,\left[1 + \frac{b^{2}\,V^{4}}{G^{2}\,(M+m)^{2}} \right]^{-1}\\
&= \frac{2\,m\,{\bf V}}{(M+m)\,(1+\alpha^{2})}
\end{split}
\eeq
where ${\bf V} \equiv {\bf v}_{m}-{\bf v}_{M}$ (i.e. the velocity of $m$ in the rest frame of $M$), $b$ is the impact parameter, and $\alpha \equiv b\,V^{2}/G\,(M+m)$ parameterizes the encounter strength. Note that the perpendicular deflection $\Delta {\bf v}_{\bot}$ will be cancelled by symmetry if the medium is homogeneous and isotropic so we neglect it for now, but we will return to this below. To account for the contributions of \textit{all} encounters, C43 then integrates Eq. \ref{eqn:one_encounter}, by noting that the encounter rate in a differential time $d t$ is the sum of encounters within a cylindrical slice, with surface area $dA$ in the plane perpendicular to the relative motion and height $V\,dt$, over all relatively velocities and angles
\beq
\begin{split}
\label{eqn:integral} {\bf a}_{\rm df} \equiv \frac{d {\bf v}_{M}}{d t} 
&= \int \Delta {\bf v}_{\|} \,V\,dA\,\mathcal{N}({\bf x},\,{\bf v})\,d^{3}{\bf v}
\\
&= \int \frac{2\,\alpha}{1+\alpha^{2}}\,\frac{G\,m}{b}\,\hat{\bf V}\,\mathcal{N}({\bf x},\,{\bf v})\,dp\,dq\,d^{3}{\bf v}
\end{split}
\eeq
where $\mathcal{N}({\bf x},\,{\bf v})=dN/d^{3}{\bf x}\,d^{3}\,{\bf v}$ is the phase space distribution function (by number) of the background particles; $\hat{\bf V} \equiv {\bf V}/V$, and $p$ and $q$ are the two spatial coordinates perpendicular to the path length $ds$, i.e. characterizing the surface $dA$ (so $ds\,dp\,dq=d^3{\bf x}$). The integral can be easily carried out for an isotropic and homogeneous distribution with $\mathcal{N}({\bf x},\,{\bf v})=nf_\text{M}(\bf{v})$, where $n$ is the number density (constant) and $f_\text{M}(\bf{v})$ is the Maxwellian velocity distribution, leading to the classical formula.

To generalize the above formula to an arbitrary phase space distribution sampled by a discrete set of data points as in our simulations, one might naively attempt to directly insert the usual $N$-body approximation, replacing $\mathcal{N}({\bf x},\,{\bf v}) \rightarrow \sum_{i} (\Delta m_{i}/m)\,\delta({\bf x}-{\bf x}_{i},\ {\bf v}-{\bf v}_{i})$. This treats the distribution function as a sum of Dirac $\delta$-functions, i.e.\ point particles, each with $N$-body particle mass $\Delta m_{i}$, so representing $N=\Delta m_{i}/m$ ``background'' particles of mass $m$. However, the integral in Eq.~\ref{eqn:integral} only integrates over the two-dimensional surface ($d p d q$) as a slice of the full phase space, which makes it impossible to discretize directly. The missing integral parameter reflects the fundamental conceptual difficulty in deriving the DF formula for \textit{arbitrary} phase space distribution. In deriving Eq.~\ref{eqn:integral}, we actually already performed the integral over the missing degree of freedom when calculating $\Delta {\bf v}_{\|}$, by integrating over path length $ds$ in each encounter from $-\infty$ to $\infty$, containing the full effect of one two-body encounter before we sum them up to get the final result. This is only correct if the background distribution is isotropic and homogeneous, since in principal, the DF process cannot be evaluated in this manner for any given instant of time, without knowing all the history and future of the full dynamics, unless the background profile is static (i.e. isotropic and homogeneous). Nevertheless, it is still suggestive to consider what an inhomogeneous background particle distribution could bring (quantitatively) to this story, hence we offer an ad hoc derivation here. 

The key conceptual requirement to replace Eq.~\ref{eqn:integral} with one that can be discretized for an arbitrary $\mathcal{N}$ is to re-expand the integral that gave rise to $\Delta {\bf v}_{\|}$ (Eq.~\ref{eqn:one_encounter}) to explicitly account for the contributions of particles at different distances $s$ along their two-body encounter trajectory, i.e. taking $\Delta {\bf v}_{\|} \rightarrow \int \langle d\Delta {\bf v}_{\|}/ds \rangle_{\rm deflected}\,ds$ (see comparison in Fig. \ref{fig:derivation}). Recall that the entire point of our derivation is to develop a formula which can be applied where the {\em explicit} $N$-body evolution of the mass $M$ was not followed. Since DF fundamentally arises from the ``back-reaction'' of the medium (i.e.\ the deflection of mass $m$ as it feels gravity from $M$ creating a net ``drag''), we need to identify the {\em difference} between the contribution to $d {\bf v}_{M}/dt$ which $m$ {\em would have} at a distance $r$ along its encounter trajectory with $M$ if it had indeed been deflected by $M$, relative to the acceleration $M$ would feel if it saw $m$ on an ``un-deflected'' trajectory. The latter is, of course, just the ``normal'' gravitational acceleration on $M$.\footnote{This contribution will differ depending on the sign of $s$ at a given $r$, i.e.\ depending on whether $m$ is ``approaching'' or ``receding'' from $M$; however in our application to $N$-body simulations, the sign of ${\bf V}$ for distant $m$ will change frequently, so there is no way to unique identify ``approaching'' or ``receding'' elements without actually performing the full time integral of every encounter (i.e.\ doing the full ``live'' $N$-body calculation with $M$, exactly what we wish to avoid). We therefore simply average between the two, giving $\langle d\Delta {\bf v}_{\|}/ds \rangle_{\rm deflected} \equiv (1/2\,|ds|)\,[\int_{-s-ds}^{-s}\,({\bf a}^{\prime}-{\bf a}^{0})\,dt+\int_{s}^{s+ds}\,({\bf a}^{\prime}-{\bf a}^{0})\,dt]$, where ${\bf a}^{\prime}\equiv {\bf a}_{Mm} [ {\bf x}_{M}^{\rm deflected}(t),\,{\bf x}_{m}^{\rm deflected}(t) ]$ and ${\bf a}^{0} \equiv  {\bf a}_{Mm} [{\bf x}_{M}^{m-{\rm undeflected}}(t),\,{\bf x}_{m}^{\rm undeflected}(t)]$ are the two-body accelerations assuming $m$ follows the deflected and un-deflected trajectories, respectively (note $M$ still ``sees'' $m$ in its un-deflected trajectory, but $m$ does not ``see'' $M$ in that case)}
The full expressions for this are quite cumbersome and cannot be analytically closed; but they are still, in any case, approximate (as we still ignore many effects such as other influences on the orbit of $m$ during each stage of its 2-body encounter), so we can safely approximate them to the same order of accuracy by noting that asymptotically $\langle d\Delta {\bf v}_{\|}/ds \rangle_{\rm deflected} \rightarrow \Delta {\bf v}_{\|}\,b^{2}/2\,(s^{2}+b^{2})^{3/2}$ at large $r\gg b$ (noting $r^{2}\equiv s^{2}+b^{2}$), and (for weak encounters, the only case where our derivation is meaningful) near pericenter ($r=b\,(1+\epsilon)$ with $\epsilon \ll 1$) $\langle d\Delta {\bf v}_{\|}/ds \rangle_{\rm deflected} \rightarrow \Delta {\bf v}_{\|}\,(1/2\,b)$. Together with the identity $1 = (b/2)\,\int_{-\infty}^{+\infty}\,b\,ds/(s^{2}+b^{2})^{3/2}$, we can replace $\Delta {\bf v}_{\|}$ in Eq.~\ref{eqn:integral} with this expression, giving:
\begin{align}
\nonumber {\bf a}_{\rm df}
&= \int \frac{2\,\alpha\,G\,m\,\mathcal{N}({\bf x},\,{\bf v})}{b\,(1+\alpha^{2})}\,\hat{\bf V}\,dp\,dq\,d^{3}{\bf v}\,\frac{b}{2}\int_{s} \frac{b\,ds}{(s^{2}+b^{2})^{3/2}}
\\
\nonumber &\approx \int \int_{s} \frac{2\,\alpha\,G\,m\,\mathcal{N}({\bf x},\,{\bf v})}{b\,(1+\alpha^{2})}\,\hat{\bf V}\,dp\,dq\,d^{3}{\bf v}\,\frac{b}{2} \frac{b\,ds}{(s^{2}+b^{2})^{3/2}}\\
\label{eqn:analytic.arbitrarydf} &= \int \frac{\alpha\,b\,G\,m}{(1+\alpha^{2})\,r^{3}}\,\hat{\bf V}\,\mathcal{N}({\bf x},\,{\bf v})\,d^{3}{\bf x}\,d^{3}{\bf v}
\end{align}
where we used $ds\,dp\,dq \equiv d^{3}{\bf x}$, and in the $\approx$ step, where we move the integrand, we essentially make a much weaker version of the original \citet{chandrasekhar1943} approximation, assuming that quantities such as $\mathcal{N}$ do not vary strongly over the timescale during which most of the $\Delta {\bf v}_{\|}$ is imparted by each 2-body encounter. Now, we can insert the discrete $N$-body form of $\mathcal{N}$ as a sum of $\delta$ functions to trivially obtain:
\beq
\label{eqn:discrete_df}
\begin{split}
{\bf a}_{\rm df}
&\rightarrow \sum_{i} \frac{\alpha_{i}\,b_{i}\,G\,\Delta m_{i}}{(1+\alpha_{i}^{2})\,r_{i}^{3}}\,\hat{\bf V}_{i}
\\
&= \sum_{i} \left( \frac{\alpha_{i}}{1+\alpha_{i}^{2}} \right)\,\left( \frac{b_{i}}{r_{i}} \right)\,\left( \frac{G\,\Delta m_{i}}{r_{i}^{2}} \right)\,\hat{\bf V}_{i}
\end{split}
\eeq

We have of course made a number of assumptions to derive Eq.~\ref{eqn:discrete_df}, and our final expression is not necessarily unique. However it has many useful properties. (1) In a spatially homogeneous medium (i.e.\ any where we can write $\mathcal{N} = n\,f({\bf v})$), then it is trivial to verify by inserting this in Eq.~\ref{eqn:analytic.arbitrarydf} that Eq.~\ref{eqn:discrete_df} reproduces {\em exactly} the expressions from \citet{chandrasekhar1943} for any $f({\bf v})$. (2) Eq.~\ref{eqn:discrete_df}, as intended, can be easily applied to an arbitrary $N$-body simulation collection of particles of arbitrary types (summing different components such as dark matter, gas, or stars simply involves carrying out the sum in Eq.~\ref{eqn:discrete_df} with the appropriate $\Delta m_{i}$ and $m$ for each ``species''). (3) Eq.~\ref{eqn:discrete_df} removes a number of ambiguities: the Coulomb logarithm is removed (it only ``re-appears'' if indeed the medium is infinite and homogeneous), and the ${\bf V}$ which appears is un-ambiguous (discussed further below). (4) Eq.~\ref{eqn:discrete_df} above can be trivially generalized for softened gravity (below). (5) Eq.~\ref{eqn:discrete_df} at least asymptotically captures the relative contributions of near versus far particles $m$ to the DF force, i.e.\ the dimensional scaling with $r$, e.g.\ correctly capturing the fact that most of the effect comes from when particles are near-pericenter.

\subsection{Force Softening}
\label{appendix:softening}

To apply Eq.~\ref{eqn:discrete_df} to numerical simulations, we must account for force softening as in the simulations (since an $N$-body particle of mass $\Delta m_{i}$ represents many individual stars, collocating them at a specific ${\bf x}_{i}$, ${\bf v}_{i}$ would lead to spurious divergences in the forces). In Eq.~\ref{eqn:discrete_df}, note that all but one term are well-behaved: $0 < \alpha_{i}/(1+\alpha_{i}^{2}) < 1/2$, $0 < b_{i}/r_{i} < 1$, and $|\hat{\bf V}|=1$, so numerical divergence entirely arises from the term $G\,\Delta m_{i}/r_{i}^{2}$. But this is just the Newtonian gravity from a point $N$-body particle -- i.e.\ {\em exactly} the same term that is force-softened in the simulations. Hence we insert the same softening kernel $S_{i}(r_{i})$ as used in the actual $N$-body simulation (taking $G\,\Delta m_{i}/r_{i}^{2} \rightarrow S_{i}(r_{i})\,G\,\Delta m_{i}/r_{i}^{2}$).

For the specific simulations here, this follows from the adaptive gravitational softening scheme described in \cite{gizmo}, corresponding to a cubic spline mass distribution:
\beq
\label{eqn:force.softening}S_{i}(r_i)= \begin{cases}
\frac{32}{3}q_i^3-\frac{192}{5}q_i^5+32q_i^6\qquad& 0\leq q_i<\frac12 \\
-\frac{1}{15}+\frac{64}{3}q_i^3-48q_i^4\\
\qquad+\frac{192}{5}q_i^5-\frac{32}{3}q_i^6\qquad& \frac12\leq q_i<1 \\
1\qquad& q_i\geq1 \\
\end{cases}
\eeq
where $q_{i}\equiv r_{i}/H_{i}$ with $H_{i}\approx 2.8\,\epsilon_{i}$ the radius of compact support of the kernel and $\epsilon_{i}$ the equivalent Plummer softening. This removes the numerical divergence and gives the correct result for a uniform density distribution sampled by $N$-body particles.
\footnote{Note that in principle this softening is not exactly self-consistent with our derivation, since if $\Delta m_{i}$ represents an extended spatial distribution of particles, each would be deflected slightly differently in Eq.~\ref{eqn:analytic.arbitrarydf}. However, this {\em is} consistent with the simulations: $N$-body softening for collisionless fluids simply features this ambiguity at a fundamental level, because an individual $N$-body particle cannot actually deform in a fully-Lagrangian manner.}

\subsection{Perpendicular Force}
\label{sec:df.perpendicular}

In the above, we only included the parallel DF term ($\propto \hat{\bf V}_{i}$). However two-body encounters also produce a perpendicular deflection ${\bf a}_{\rm df,\,\bot}$; this only vanishes in the C43 derivation because of the assumption of a homogeneous $\mathcal{N}$ (giving exact cancellation). Because we do not assume homogeneous $\mathcal{N}$, we can (if desired) retain these terms, giving:
\begin{align}
\label{eqn:df.perp}
    {\bf a}_{\rm df,\,\bot} &= -
    \sum_{i} \left(\frac{1}{1+\alpha_{i}^{2}} \right)\,
    \left( \frac{b_{i}}{r_{i}} \right) \,
    \left( S_{i}(r_{i})\,\frac{G\,\Delta m_{i}}{r_{i}^{2}} \right)\,
    \hat{\bf b}_{i} \\ 
\nonumber {\bf b}_{i} &\equiv {\bf r}_{i} - ({\bf r}_{i}\cdot \hat{\bf V}_{i})\,\hat{\bf V}_{i}
\end{align}
This differs from the parallel ${\bf a}_{\rm df,\,\|}$ only by one power of $\alpha_{i}$ and, of course, the direction. The power of $\alpha_{i}$ means that the perpendicular deflection can be stronger (compared to the parallel term) in strong encounters (although $0<1/(1+\alpha_{i}^{2})<1$ so this term is still bounded and cannot produce spurious divergences or forces larger than the regular/external acceleration). But because the integrated force is always dominated by weak deflections (where $\alpha_{i} \gg 1$), then even ignoring cancellations (which further reduce ${\bf a}_{\rm df,\,\bot}$ even in inhomogeneous $\mathcal{N}$), this term is generally smaller than the parallel $|{\bf a}_{\rm df,\,\|}|$ by one power of $\sim G\,M/r\,V^{2} \sim M/M_{\rm total,\,galaxy}(<r) \ll 1$.

We show in an additional set of tests that this term is completely negligible for most galaxy simulation contexts, hence we do not include them in our final expression and tests below. But we emphasize that it is trivial to include and imposes no additional cost.

\subsection{Final Expression}
\label{appendix:expression}

It is straightforward to generalize the above for a spectrum of masses $m$, i.e.\ integrating over the stellar initial mass function (IMF). However for any $M \gtrsim 10\,M_{\odot}$, this makes a negligible difference to our results. Since we do not know the ``true'' dark matter particle mass, it is more straightforward to simply assume the limit $M \gg m$, in which case the species masses $m$ completely factor out of the salient expressions. This gives the expression we will use throughout:
\begin{equation}
\begin{split}
\label{eqn:discrete_df_smoothing}
     {\bf a}_{\rm df} &= \sum_{i} \Delta {\bf a}_{\rm df}^{i} \\ 
    \Delta {\bf a}_{\rm df}^{i} &\equiv 
    \left( \frac{\alpha_{i}\,b_{i}}{(1+\alpha_{i}^{2})\,r_{i}} \right)\,
    \left( S_{i}(r_{i})\,\frac{G\,\Delta m_{i}}{r_{i}^{2}} \right)\,
    \hat{\bf V}_{i}
\end{split}
\end{equation}
with $\alpha_{i} \approx b_{i}\,V_{i}^{2}/G\,M$.

\subsection{Numerical Implementation}

In the form of Eq.~\ref{eqn:discrete_df_smoothing}, it is particularly straightforward to implement our estimator. First, noting that $\alpha_{i}$ and $b_{i} \equiv r_{i}\,|\hat{\bf r}_{i} - (\hat{\bf r}_{i}\cdot \hat{\bf V}_{i})\,\hat{\bf V}_{i}|$ is a function only of ${\bf r}_{i}$ and ${\bf V}_{i}$, we see that the only piece of additional of information needed to compute Eq.~\ref{eqn:discrete_df_smoothing}, alongside the usual gravity force, in an $N$-body solver is the velocity ${\bf V}$ (already known). In other words, we do not need to construct some estimator for values in the C43 formula, like $\rho$, $\Lambda$, $\langle {\bf V} \rangle$ which are not actually computed in standard $N$-body simulations. Second, we also immediately see that it is completely trivial to carry out this sum over any arbitrary set of species (e.g.\ stars+gas+dark matter+other BHs).

Comparing the form of Eq.~\ref{eqn:discrete_df_smoothing} and the ``regular'' gravitational acceleration ${\bf a}_{\rm ext}$:
\begin{align}
{\bf a}_{M} &= {\bf a}_{\rm ext} + {\bf a}_{\rm df} = \sum_{i}\Delta {\bf a}_{\rm ext}^{i} + \sum_{i}\Delta {\bf a}_{\rm df}^{i} \\ 
\Delta {\bf a}_{\rm ext}^{i} &\equiv \left( S_{i}(r_{i})\,\frac{G\,\Delta m_{i}}{r_{i}^{2}} \right)\, \hat{\bf r}_{i} \\ 
    \Delta {\bf a}_{\rm df}^{i} &\equiv 
    \left( \frac{\alpha_{i}\,b_{i}}{(1+\alpha_{i}^{2})\,r_{i}} \right)\,
    \left( S_{i}(r_{i})\,\frac{G\,\Delta m_{i}}{r_{i}^{2}} \right)\,
    \hat{\bf V}_{i}
\end{align}
we immediately see that the operation needed to compute ${\bf a}_{\rm df}$ is algorithmically identical to that needed to compute the normal gravitational forces. In Tree-gravity, Tree-PM, direct $N$-body, or many other methods, implementing exact evaluation of Eq.~\ref{eqn:discrete_df_smoothing} in a manifestly-conservative manner is especially trivial.
\footnote{In PM and related methods, where long-range forces are evaluated via computing the potential from a Particle-Mesh Fourier method, implementing Eq.~\ref{eqn:discrete_df_smoothing} is less trivial: the issue is that the direction $\hat{\bf V}_{i}$ differs from $\hat{\bf r}_{i}$, so one cannot simply treat ${\bf a}_{\rm sf}$ as a scalar correction to the regular external gravitational potential, but must compute a separate potential/field. However in hybrid Tree-PM methods, such as (optionally) implemented in {\small GIZMO}, the less-accurate PM forces are only used at large distances; given this, we find (consistent with Fig.~\ref{fig:lss_slice}) that the errors from simply truncating the sum for ${\bf a}_{\rm df}$ by including only the contributions from the tree-walk (ignoring the PM terms in ${\bf a}_{\rm df}$) are entirely negligible (below normal integration-error level).} 
In e.g.\ a tree-walk, as one sums up to compute ${\bf a}_{\rm ext}$, we simply sum the additional term $\Delta {\bf a}_{\rm df}^{i}$, which scales exactly with the $| \Delta {\bf a}_{\rm ext}^{i} | $ multiplied by the numerical pre-factor $\alpha_{i}\,b_{i}/(1+\alpha_{i}^{2})\,r_{i}$, and oriented in the different direction $\hat{\bf V}_{i}$. The gravitational force softening is also naturally embedded in Eq. \ref{eqn:discrete_df_smoothing}.

Moreover, our Eq.~\ref{eqn:discrete_df_smoothing} is well-behaved when applied to tree nodes/leaves, not just individual particles: one simply treats each node as a ``super-particle'' with the appropriate total $\Delta m_{i}$ and mass-averaged ${\bf V}_{i}$, ${\bf r}_{i}$, in the same manner as done for the usual gravity calculation. It is trivial to verify from the form of Eq.~\ref{eqn:discrete_df_smoothing} that the order of the errors from this approach will always be equal to or better than the order of errors in ${\bf a}_{\rm ext}$ in the tree (i.e.\ convergence is equal or faster).

To ensure manifest momentum conservation, we simply enforce equal and opposite forces, i.e.\ apply an acceleration $\Delta {\bf a}_{\rm M-to-i} = -(M/\Delta m_{i})\,\Delta {\bf a}_{\rm df}^{i}$ to each particle $i$. The scaling of the pre-factor in Eq.~\ref{eqn:discrete_df_smoothing} is such that it guarantees this ``back-reaction'' term is well behaved and does not produce any spurious numerical divergences in the accelerations of the particles $i$.\footnote{That behavior is {\em not} guaranteed if one attempts to conserve momentum by simply applying a C43-style formula to $M$ and then ad-hoc ``redistribute'' the equal-and-opposite momentum change to the neighboring $i$ around $M$.}

\section{Numerical Validation: Methodology} 
\label{sec:methods}

To study the accuracy of our DF formula, we compare it to both direct high-resolution simulations and calibrated versions of the local Chandrasekhar's DF formula, using both ``on-the-fly'' applications in simulations (\S~\ref{sec:methods:sim}) and post processing methods (\S~\ref{sec:methods:post}). Here we detail those methods. In what follows, we refer to the ``target'' or ``sinking'' particle as a black hole (BH) of mass $M_{\rm BH}$, since this is a particularly relevant motivating case for our sub-grid model, but of course the ``target'' particle could in principle represent any sufficiently compact bound massive object. 

\subsection{On-the-fly Simulations}
\label{sec:methods:sim}

\subsubsection{Numerical Methods}

We have implemented the ``discrete DF estimator'' Eq.~\ref{eqn:discrete_df_smoothing} in the {\small GIZMO} multi-physics code \citep{gizmo}, which uses a standard Barnes-Hut tree algorithm to solve the gravity equations \citep[an improved version of that in][]{springel:gadget}. {\small GIZMO} is well-tested in numerous applications of $N$-body dynamics problems involving dynamical friction, $N$-body resonances and wake problems (see, e.g. \citealt{Lokas2019, Collier2020, Grudic2020, bonetti2020, Morton2021, Bonetti2021, Bortolas2022}), to which we refer for more detailed descriptions of numerical methods, demonstrations of convergence, test problems, etc. As described above we simply evaluate the DF force ${\bf a}_{\rm df}$ alongside the ``normal'' gravitational force (using the identical softening, etc.) in the tree-walk operation, imposing negligible CPU cost.

\subsubsection{Initial Conditions}

To test the estimator, we have run a series of test problems. In each, we initialize a steady-state ``halo'' of collisionless particles (e.g.\ ``dark matter'' or ``stars'') using the {\small GALIC} code \citep{galic}, with a target/BH particle on an initial orbit expected to decay owing to DF. We have experimented with several different choices for the initial halo density profile, whether the halo velocity distribution is anisotropic or isotropic, and other parameters of the halo and orbit (e.g.\ eccentric versus circular, and initial position/energy/angular momentum). Our qualitative conclusions and comparison of methods are identical in each case (and of course, this being a pure $N$-body problem it is scale free), so we focus on and show plots from one example with typical cosmological units for the sake of clarity. 

In our fiducial example, we adopt a \citet{hernquist1990}-profile halo with total mass $2\times10^{11}\,M_\odot$ with the \citet{galic} concentration parameter of $4$ and spin parameter $0.04$ (consistent with typical dark matter halo parameters, \citealt{Bullock2001}, and sufficient to make the halo mildly anisotropic because of rotation), so that the \citet{hernquist1990} scale-length $a=30.2\,{\rm kpc}$. The target/BH is placed $5\,\mathrm{kpc}$ away from the halo center and has a tangential velocity of $59\,\mathrm{km/s}$, which is the circular velocity of the halo at that radius. The black hole mass is $10^8\,M_\odot$, much less than the enclosed dark matter mass inside $5\,\mathrm{kpc}$ ($\sim 4\times10^9\,M_\odot$), to avoid disrupting the dynamical equilibrium of the galaxy.

\subsubsection{Sub-Grid Versus Resolved Simulations}

As DF should be fully resolved when the target/BH mass $M_\mathrm{BH}$ is much more than the background (``dark matter'' or DM) particle mass $M_\mathrm{DM}$, one would expect that only in a low resolution simulation (i.e., $M_\mathrm{BH}\lesssim M_\mathrm{DM}$) a sub-grid treatment of DF is necessary\footnote{
When the resolution lies in between and DF is partially resolved, a sub-grid treatment may cause ``double counting'' when calculating DF. While this remains an open question in general, we find that it can be avoided by multiplying a field-mass-dependent function on the DF formula in our tests. See discussions in \ref{sec:double-count}.}. However, if the resolution is {\em too} low, the orbital semi-major axis of the BH particle will be smaller than the inter-particle spacing of the $N$-body simulation and the BH will have essentially ``sunk to the center'' already -- trivially, if it were just one background/DM particle inside of the initial $5\,\mathrm{kpc}$, then there is no define-able smaller-scale center towards which even a ``perfect'' sub-grid model could migrate the target/BH. We hence choose $M_\mathrm{DM}=10^7M_\odot$ in the tests with sub-grid DF, so $\sim 400$ dark matter particles are enclosed inside the initial $5\,\mathrm{kpc}$. We further run a set of 50 simulations with the same background halo, but with the BH particles placed randomly on a $5\,\mathrm{kpc}$-radius sphere with a random direction of velocity in the tangent plane. By choosing the median between these runs, we can smooth out the chaotic motions intrinsic in the problem, as well as the effects of anisotropy (both real, from the halo rotation, and numerical, from $N$-body noise) generating eccentric orbits which produce larger oscillations in the instantaneous BH speed (making the results more difficult to read).

To test our results, we compare a set of reference simulations at varying resolution which do not adopt any sub-grid DF, but with the same setups of black hole initial conditions. At sufficiently high resolution, these simulations satisfy $M_\mathrm{BH} \gg M_\mathrm{DM}$ and so should directly capture the salient effects of DF on the target.

\subsubsection{Simulations with a ``Fitted'' C43 Sub-Grid Model}

Finally, we consider a third set of simulations where we again adopt a sub-grid DF estimator, but instead adopt the local Chandrasekhar DF estimator of Eq.~\ref{eqn:chandra_maxwell} as previously introduced in {\small GADGET} in e.g.\ \citet{cox:xray.gas} updated to be essentially identical to that in \citet{Tremmel2015}. Here we assume a Maxwellian velocity distribution, estimate the mean velocity and dispersion as a kernel-and-cell-mass-weighted mean, and use the BH kernel density estimator from \citet{wellons:2022.smbh.growth} to estimate $\rho$. 

We previously noted intrinsic difficulties this method faces: however, for this particular test problem, the background halo is (by construction) smooth and nearly isotropic and single-component and nearly-Maxwellian, so this provides a ``best-case scenario'' for a C43-like estimator. But this still leaves un-resolved the question of how to estimate the Coulomb logarithm. We find that common choices (e.g. the ratio of virial radius to ``true'' inter-particle spacing) are not only impossible to predict a-priori in a completely general simulation (they must be put in ``by-hand''), but also appear to give DF forces which differ systematically from the resolved solutions by tens of percent or up to a factor of two. Therefore, to give this model the best possible chance, we explicitly {\em fit} the Coulomb logarithm, varying it until we find a model which best matches the BH orbital decay seen in the explicit high-resolution $N$-body calculation. We use this, essentially as a way of detecting how our method compares to a ``best-case'' C43 estimator calibrated ahead of time to the {\em specific} problem being simulated.

The simulation setups are summarized in Table \ref{tab:simulations}.

\begin{table}
    \centering
    \begin{tabular}{ccccc}
         \\\hline
         set & $M_\mathrm{DM}/M_\mathrm{BH}$ & sub-grid DF model & $\epsilon$ criterion & num. of runs\\
         \hline
         1 & $10^{-1}$ & this paper & $\epsilon\sim\Delta x_i$ & 50\\
         2 & $10^{-1}$ & fitted C43 & $\epsilon\sim\Delta x_i$ & 50\\
         3 & $10^{0}$ & none & $\epsilon<b_\mathrm{min}$ & 50\\
         4 & $10^{-1}$ & none & $\epsilon<b_\mathrm{min}$ & 50\\
         5 & $10^{-2}$ & none & $\epsilon<b_\mathrm{min}$ & 50\\
         6 & $10^{-3}$ & none & $\epsilon<b_\mathrm{min}$ & 1\\
         7 & $10^{-4}$ & none & $\epsilon<b_\mathrm{min}$ & 1\\
         8 & $10^{0}$ & none & $\epsilon\sim\Delta x_i$ & 50\\
         9 & $10^{-1}$ & none & $\epsilon\sim\Delta x_i$ & 50\\
         10 & $10^{-2}$ & none & $\epsilon\sim\Delta x_i$ & 50\\
         11 & $10^{-3}$ & none & $\epsilon\sim\Delta x_i$ & 1\\
         12 & $10^{-4}$ & none & $\epsilon\sim\Delta x_i$ & 1\\
         13 & $10^{-5}$ & none & $\epsilon\sim\Delta x_i$ & 1\\
         \hline
    \end{tabular}
    \caption{Representative simulation summary for our idealized tests. Different sets share the same setup of initial conditions: a $10^8\,M_\odot$ black hole particle placed randomly at a $5\,\mathrm{kpc}$ radius with a velocity of $59\,\mathrm{km/s}$ in a random tangent direction. The background particles form an Hernquist halo with $M_{\rm halo}=2\times10^{11}\,M_\odot$. The black hole speed from these tests are shown in Fig. \ref{fig:values_evo} (when multiple runs are at present, only the median value is shown).}
    \label{tab:simulations}
\end{table}

\subsection{Post-Processing in Multi-Physics Galaxy Simulations}
\label{sec:methods:post}

While comparing our discrete estimator with the Chandrasekhar estimator in the idealized test problem above can help to test its accuracy, it is of course also important to apply it to some more ``realistic'' (or at least more complicated) galaxy simulations which involve multi-component (gas+star+DM) anisotropic, highly-inhomogeneous backgrounds. Full applications to such simulations on-the-fly can be used to make predictions for e.g.\ demographics of free-floating BHs, IMBHs, and rates of BH-BH coalescence in galaxy nuclei (e.g. predictions for LISA). However this is clearly beyond the scope of this work. Instead here we will select some snapshots of $N$-body information from high-redshift galaxies in the Feedback In Realistic Environments (FIRE; \citealt{fire,fire2}) project, and use these to make some simple post-processing comparisons in order to see how the full on-the-fly application of the estimator used here might differ (or not) from other approaches to including or ignoring DF in these kinds of systems.

\section{Results and Discussions} \label{sec:results}
\subsection{Validation in On-The-Fly Simulations}
\label{sec:results_sim}

\begin{figure*}
    \centering
    \includegraphics[width=\textwidth]{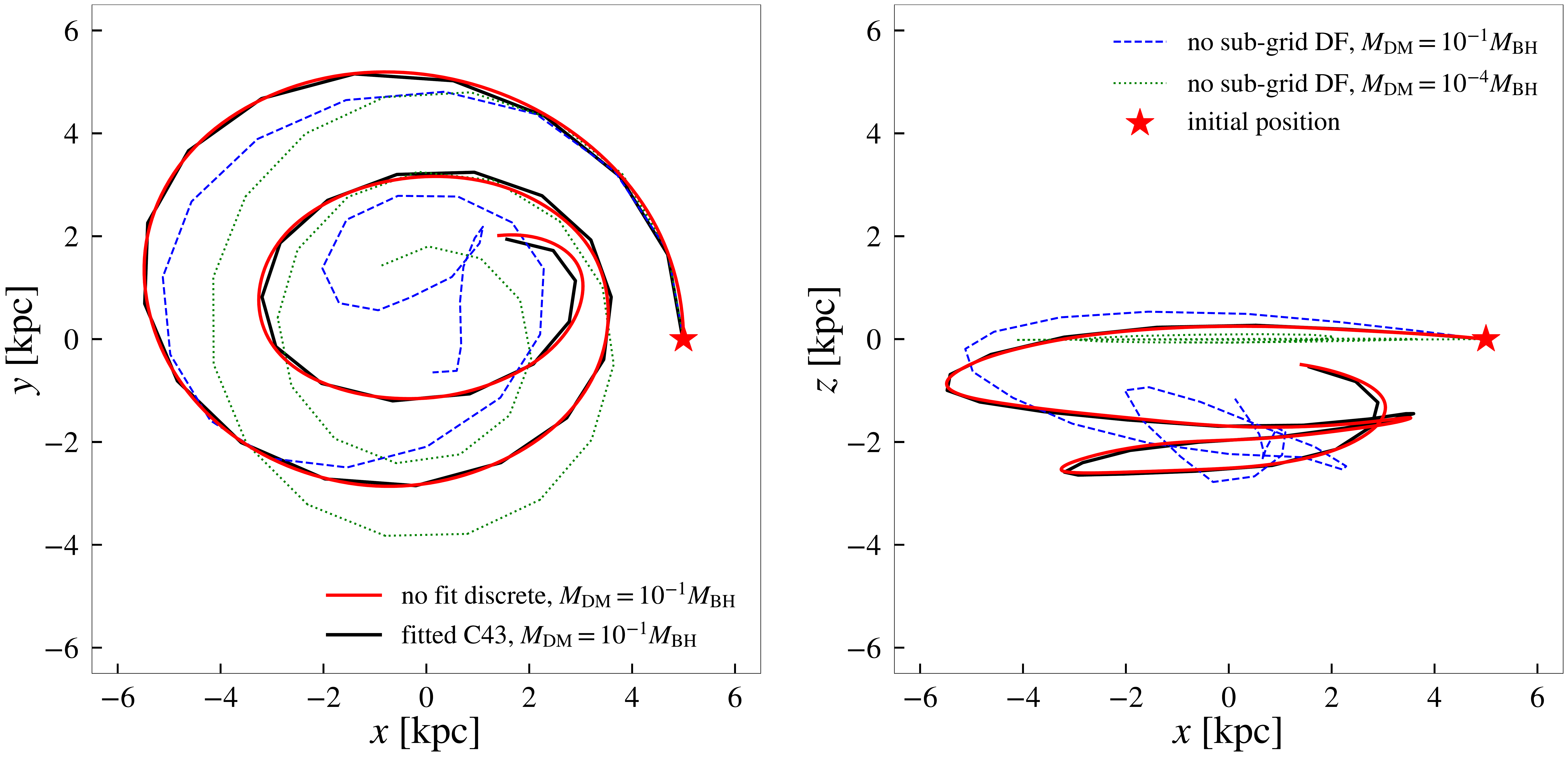}
    \caption{Example trajectories of a black hole particle in our simulations. The black hole is initially placed $5\,\mathrm{kpc}$ away from the halo center (the coordinate origin) on the $x$-axis and has a circular velocity of $59\,\mathrm{km/s}$ in the $\hat{y}$ direction. We see that in the high-resolution run (green dashed) the black hole sinks to the halo center in circular orbits as time evolves, which is partially resolved by the low-resolution runs with sub-grid DF (red and black lines, with the discrete DF and the ``calibrated Chandrasekhar'' estimator, respectively), but not by the run without it (blue dashed; the black hole departs significantly from the halo center in the $z$-direction). The low-resolution runs (with or without sub-grid DF) suffer from dynamical heating which significantly perturbs the circular orbit. In addition, our discrete estimator matches well with the calibrated Chandrasekhar estimator -- but we have not calibrated our discrete DF estimator in any way (we are simply using Eq.~\ref{eqn:discrete_df_smoothing} directly, without any input parameter other than the smoothing length $\epsilon$).}
    \label{fig:trajectories}
\end{figure*}

\begin{figure*}
    \centering
    \includegraphics[width=\textwidth]{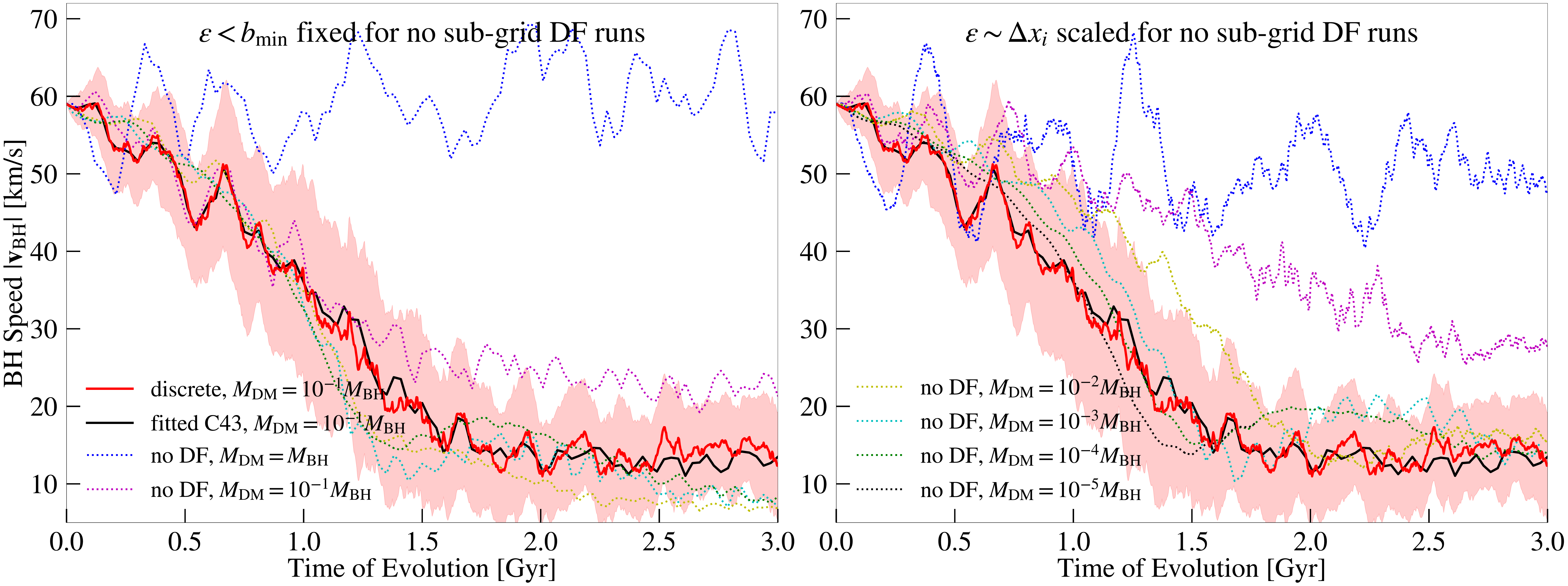}
    \caption{The speed of black hole particles upon time of evolution in our test problem. The thick red, thick black and dotted lines show the (median of) results from low-resolution runs with our discrete estimator, with Chandrasekhar's DF estimator (with a fitted $\ln\Lambda=4$) and from multi-resolution runs without sub-grid DF, respectively (see Table \ref{tab:simulations}). The red-shaded area shows the $\pm\sigma$ range for all runs with the discrete estimator. We see that the speed of black hole particles decreases significantly as the black holes sink to the halo center. The results from our discrete estimator matches well with the fitted Chandrasekhar estimator, and matches the converged results of the no sub-grid DF runs at higher resolution. The convergence is better for no sub-grid DF runs with smoothing length less than the minimum impact-parameter ($\epsilon<b_\mathrm{min}$, left panel) than those with (the usually chosen) length comparable to the inter-particle separation ($\epsilon\sim\Delta x_i$, right panel), as in the later case the effective Coulomb logarithm is artificially truncated, causing a logarithmic convergence behaviour (see discussions in \S \ref{sec:results_sim}).}
    \label{fig:values_evo}
\end{figure*}

Figs.~\ref{fig:trajectories}-\ref{fig:values_evo} show some representative results of our numerical validation tests in on-the-fly simulations, specifically focusing on an illustrative trajectory of the BHs as well as the BH velocity as a function of time.

First, we examine the behavior of pure $N$-body calculations ({\em without} sub-grid DF) as a function of resolution. Not surprisingly, when the target mass is similar to the $N$-body particles (e.g. $m_{\rm DM} \gtrsim M_{\rm BH}$), no DF is captured. Most previous studies arguing for different ``sufficient'' resolutions to capture DF refer to this regime \citep[see e.g.][]{vandenbosch:no.orbital.circularization.due.to.dyn.frict,colpi:2007.binary.in.mgrs,boylankolchin:merger.time.calibration,fire2,Pfister2019,Barausse2020,2020MNRAS.495L..12B,Ma2021}, depending on the specific problems they are choosing. In our case, at better resolution ($m_{\rm DM} \ll M_{\rm BH}$) we see DF but with an important dependence on how we treat the {\em spatial} force softening $\epsilon$. If we adopt a fixed Plummer-equivalent $\epsilon$ comparable to or smaller than the canonical minimum impact parameter for strong encounters $b_{\rm min} \sim G\,M_{\rm BH}/(2\,\sigma^{2}+V_{\rm bh}^{2})$ (here $\sim 60\,{\rm pc}$ at the initial BH position), we see excellent convergence once $m_{\rm DM} \ll 0.1\,M_{\rm BH}$ (Fig.~\ref{fig:values_evo}, left-panel). However, this is not how force softenings are typically set in $N$-body simulations which do not resolve the individual point masses: instead, to prevent spurious noise in {\em other} properties, the ``optimal'' softening is usually chosen to roughly match the inter-particle separation $\epsilon \sim \Delta x_{i} \sim (\Delta m_{i}/\rho_{i})^{1/3}$ \citep[Fig.~\ref{fig:values_evo}, right-panel;][]{merritt:1996.optimal.softening,romeo.1998:optimal.softening,athanassoula:2000.optimal.force.softening.collisionless.sims,dehnen:2001.optimal.softening,rodionov:2005.optimal.force.softening}. When we do this, we see notably worse convergence: in fact, the convergence is logarithmic in $m_{\rm DM}$, because we have $\epsilon > b_{\rm min}$, the effective Coulomb logarithm is artificially truncated (i.e.\ we artificially suppress close encounters). This is a known challenge for DF in softened gravity \citep[see e.g.][for more details and extended discussion]{karl:2015.direct.nbody.df.sims}, and it further emphasizes the importance of a sub-grid model like ours: achieving $\Delta x_{i} \ll b_{\rm min}$ requires $m_{\rm DM} \ll 10^{-5}\,M_{\rm BH}$, i.e.\ billions of $N$-body particles even for a simple, idealized halo like that here. 

We then compare our ``sub-grid'' DF model (Eq.~\ref{eqn:discrete_df_smoothing}) calculated on-the-fly to an extremely low-resolution IC with $m_{\rm DM}=0.1\,M_{\rm BH}$,\footnote{We find that the results of our sub-grid DF runs are robust and nearly independent of resolution so long as the dynamical mass of the target/BH particle is at least slightly larger (a factor of $\gtrsim2-3$) than the mass-weighted median of the ``background'' $N$-body particles. If the BH particle has mass lighter than the background, then either sub-grid DF model (C43 or Eq.~\ref{eqn:discrete_df_smoothing}) requires additional care, or else spurious $N$-body heating effects can become larger than the true DF forces. So for practical applications where one wishes to evolve the dynamics of targets with very small masses, it is useful to follow standard practice \citep{springel:multiphase,dimatteo:cosmo.bh.growth.sim.1,hopkins:lifetimes.letter} and assign a separate ``true target/BH mass'' used for the DF calculation and other physics to the $N$-body particle ``carrying'' the target/BH.} using $\epsilon \sim \Delta x_{i}$ as would be applied in typical cosmological simulations. For this low-resolution case, there is significant variation owing to different eccentric orbits and discreteness noise, so we show the median and $\pm1\sigma$ range of BH velocities. The median agrees remarkably well with the converged solution. We stress that Eq.~\ref{eqn:discrete_df_smoothing} contains {\em no other adjustable parameter} beyond the physically motivated $\epsilon$: this is an actual prediction.

Next we compare the ``fitted'' C43 model Eq.~\ref{eqn:chandra_maxwell}: as described above, in addition to the arbitrary choice of kernel estimator size and shape (which we set to the smallest size that reduces noise acceptably), we freely vary the numerical pre-factor (``effective Coulomb log'') $\ln{\Lambda}$ in Eq.~\ref{eqn:chandra_maxwell} until we find a value which best matches our high-resolution simulations. For the best-fit value, the result is strikingly similar to our Eq.~\ref{eqn:discrete_df_smoothing} (perhaps not surprising, given that we start from the same assumptions) -- but we stress that even small, $\sim 10\%$ differences in $\ln{\Lambda}$ produce significant disagreement with the high-resolution simulations. Moreover, we have considered a dozen ``standard'' estimates of $\ln{\Lambda}$ widely used in the literature \citep[see references above and][]{hashimoto03:varying.culomb.log.in.dynfric,just:2011.dyn.fric.coulomb.calib,antonini:2012.df.faster.star.contrib,Dosopoulou:2017.df.modeling}, e.g.\ $\Lambda \sim |\rho/\nabla \rho|/b_{\rm min}$, and find that {\em none} of them correctly predicts the best-fit $\Lambda$ (usually discrepant by factors $\sim1.3-2$). This probably owes at least in part to the fact that the central \citet{hernquist1990} distribution function is appreciably non-Maxwellian, as discussed in \citep{karl:2015.direct.nbody.df.sims}, so the fitted $\ln{\Lambda}$ is essentially compensating for this error (the ``${\rm erf(...)-...}$'' term in Eq.~\ref{eqn:chandra_maxwell}).

As noted above, these conclusions are robust to the parameters of the initial halo and orbit, mass profile of the halo assumed, amount of angular momentum (anisotropy in the distribution function), and other choices of the problem setup: however, we find as expected that the C43 ``effective Coulomb logarithm'' must be re-calibrated in many cases to fit high-resolution simulations. We have also tested other numerical aspects of the method including e.g.\ the tree opening criteria \citep{power:2003.nfw.models.convergence,springel:gadget}, timestep size/integration accuracy \citep{fire2,Grudic2020}, and inclusion/exclusion of the perpendicular force (Eq.~\ref{eqn:df.perp}): none of these has a significant effect \citep[consistent with previous studies, see e.g.][]{just:2011.dyn.fric.coulomb.calib,karl:2015.direct.nbody.df.sims,mikherjee:2012.fmm.dynfric.tests}. 

Given the close agreement between the discrete DF and explicitly calibrated-Chandrasekhar DF models, it is likely that more detailed differences in orbit shape we see comparing {\em either} of these models and the true, high-resolution simulation owes not to anything we can simply ``further calibrate'' (like a Coulomb logarithm), but rather to fundamental resolution effects (e.g.\ more accurately recovering the shape of the background potential itself, hence the ``correct'' elliptical orbit structure; or the treatment of subparsec-scale physics around SMBHs, a known issue as discussed in, e.g. \citealt{Rantala2017,Mannerkoski2021,Mannerkoski2022}), as well as assumptions of the Chandrasekhar-like derivation which our DF derivation also implicitly assumes. For example, the assumption of linearity (that the net effect on the BH can be approximated via the sum of many independent two-body encounters) or forward/backward asymmetry in the distribution function (implicit in a stronger assumption like homogeneity but present in a weaker form in our derivation as well).

\subsection{Post-Processing in Multi-Physics Galaxy Simulations}
\label{sec:results_post}

\begin{figure}
    \centering
    \includegraphics[width = \columnwidth]{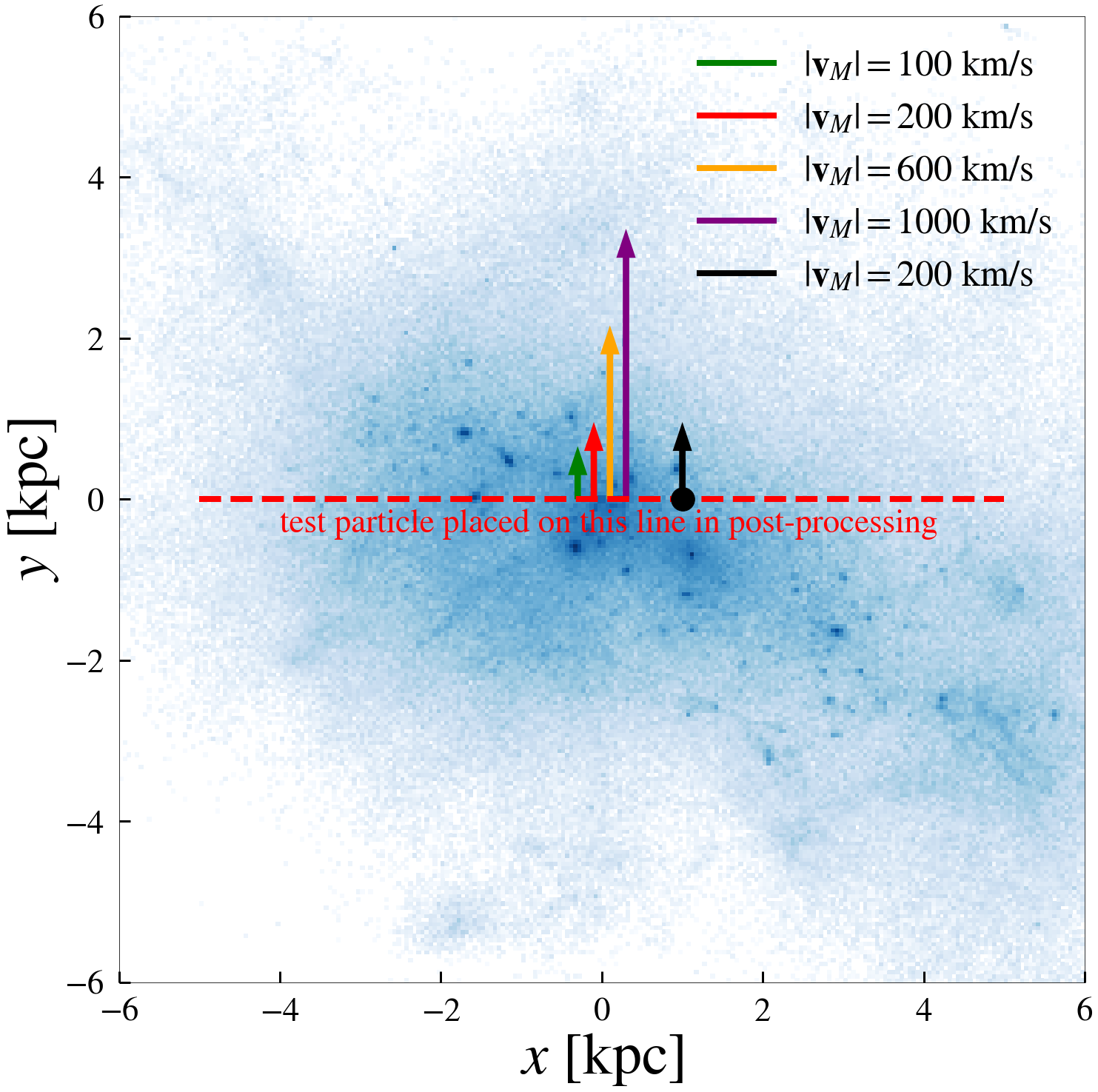}
    \caption{The galaxy snapshot we chose for post-processing analysis. The color scales with the projected total mass density (DM+stars+gas). It is the ``{\bf z5m12b}'' galaxy at redshift 7.0 described in \protect\cite{Ma2021}, which is clumpy and dynamically unstable. The post-processing tests are also shown: (a) a test particle of $10^5\,M_\odot$ placing on the $x$-axis with a velocity of $V_M=200\,\mathrm{km/s}\,\hat{y}$; (b) the same particle but with different velocities ($100, 200, 600$ and $1000\,\mathrm{km/s}$) in the $\hat{y}$ direction; (c) a fixed particle at $(1,0,0)$ with $V_M=200\,\mathrm{km/s}\,\hat{y}$. The implications of these tests are described in the main text.}
    \label{fig:galaxy}
\end{figure}

While the idealized experiments above are important for validation, their simplicity means that it is difficult to gain insight into possible differences between our Eq.~\ref{eqn:discrete_df_smoothing} and the fitted C43 model. We therefore briefly consider this in post-processing of a multi-physics galaxy formation simulation. The specific (arbitrary) simulation and time we select is the ``{\bf z5m12b}'' galaxy at redshift $7.0$ described in \cite{Ma2021}, illustrated in Fig. \ref{fig:galaxy}. The simulation is multi-component, containing dark matter, stars, multi-phase gas, and black holes, with complicated cooling, star formation and ``feedback'' physics all included on-the-fly. This particular snapshot is chosen because it is dynamically unstable, asymmetric, gas-rich and starforming, and contains several giant star clusters and molecular cloud complexes, all of which complicate the dynamics. We compare the results from the discrete estimator with the Chandrasekhar estimator and discuss their differences and implications.

\begin{figure}
    \centering
    \includegraphics[width=\columnwidth]{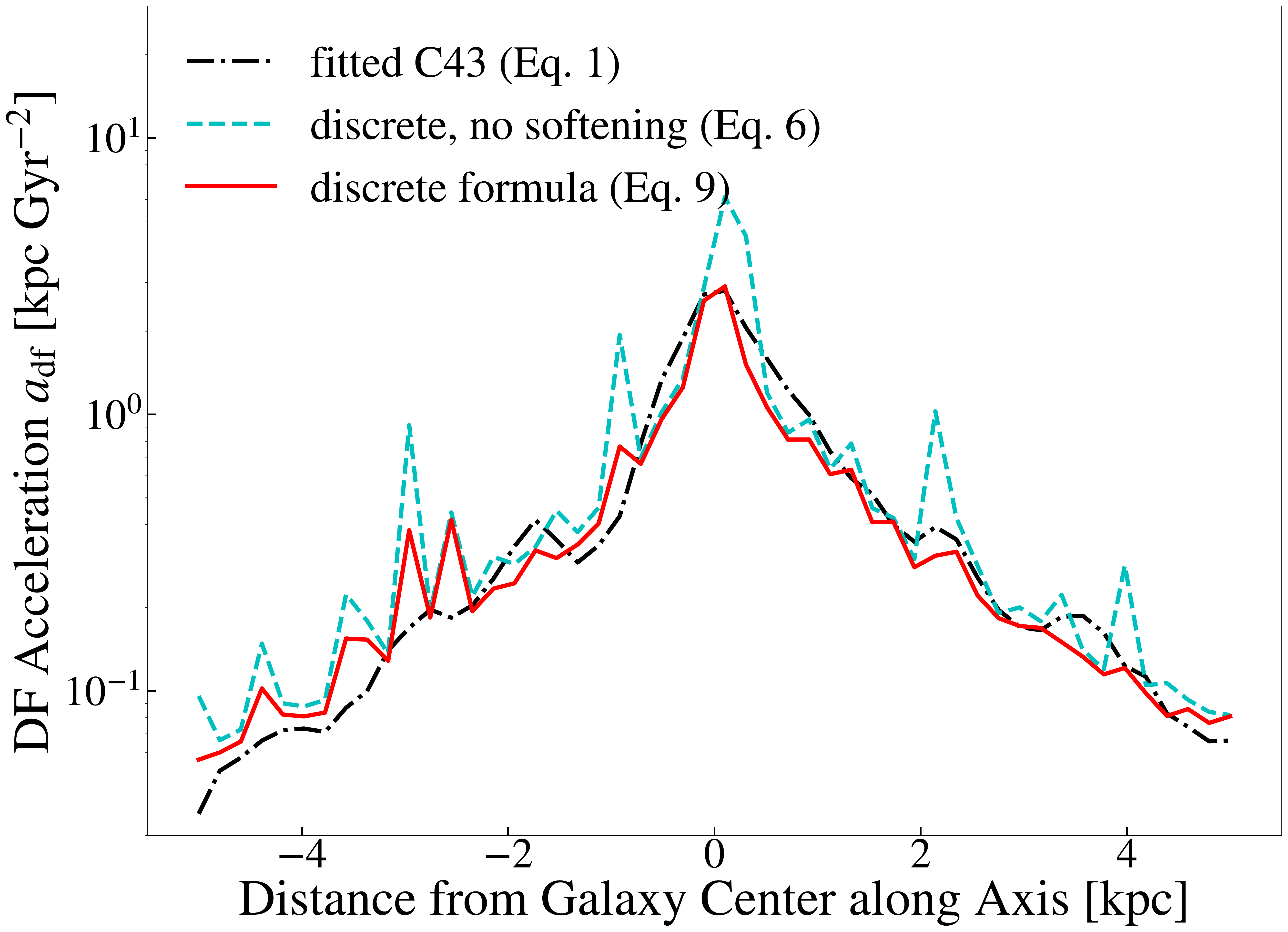}
    \caption{Comparison of the DF amplitude calculated from different DF formulas. The test particle is a $10^5 M_\odot$ particle with a $200\,\text{km/s}$ velocity in the $y$ direction, put at different positions on the $x$ axis (Fig. \ref{fig:galaxy}, red dashed line and arrow). The black, cyan and red lines show the results from Chandrasekhar's formula (Eq. \ref{eqn:chandra_original}, with a (fitted) constant Coulomb logarithm $
\ln\Lambda = 5$), our formula without smoothing (Eq. \ref{eqn:discrete_df}) and our formula with smoothing (Eq. \ref{eqn:discrete_df_smoothing}), respectively. Our discrete formula remains very close to Chandrasekhar's approximation. The smoothing removes most of the peaks which could be caused by numerical divergence.}
    \label{fig:df_compare_amp}
\end{figure}

\begin{figure}
    \centering
    \includegraphics[width=\columnwidth]{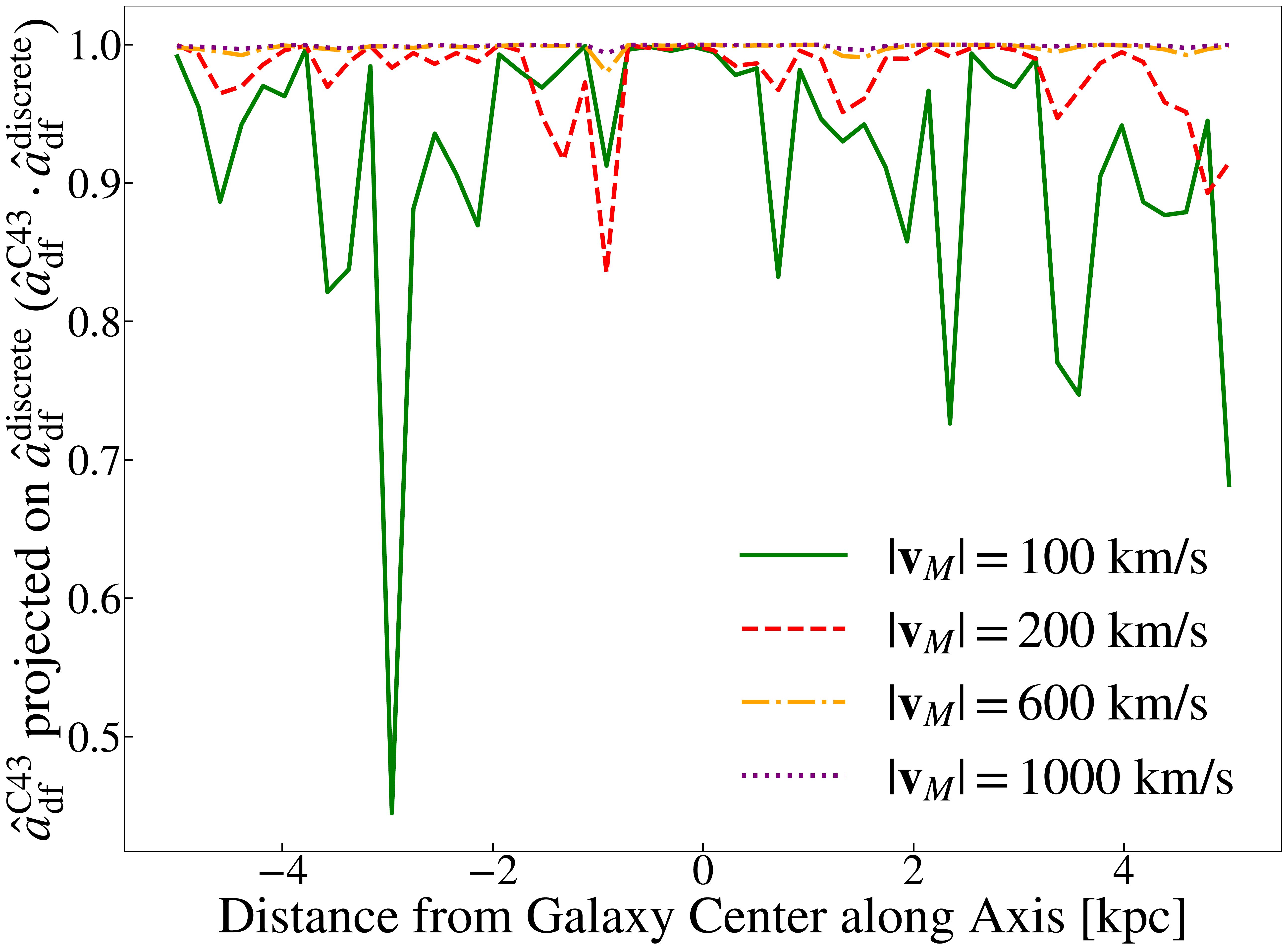}
    \caption{The unit direction vectors calculated from Chandrasekhar's projected on those from our DF formulas. The test particle is the same as in Fig. \ref{fig:df_compare_amp} but with different velocities (still along the $y$ axis). The difference of this value from 1 shows how mis-aligned the directions are. At most positions the directions are perfectly aligned for relative high velocities, yet at low velocities huge error could occur due to contributions from particles far apart.}
    \label{fig:df_compare_dir}
\end{figure}

\begin{figure}
    \centering
    \includegraphics[width=\columnwidth]{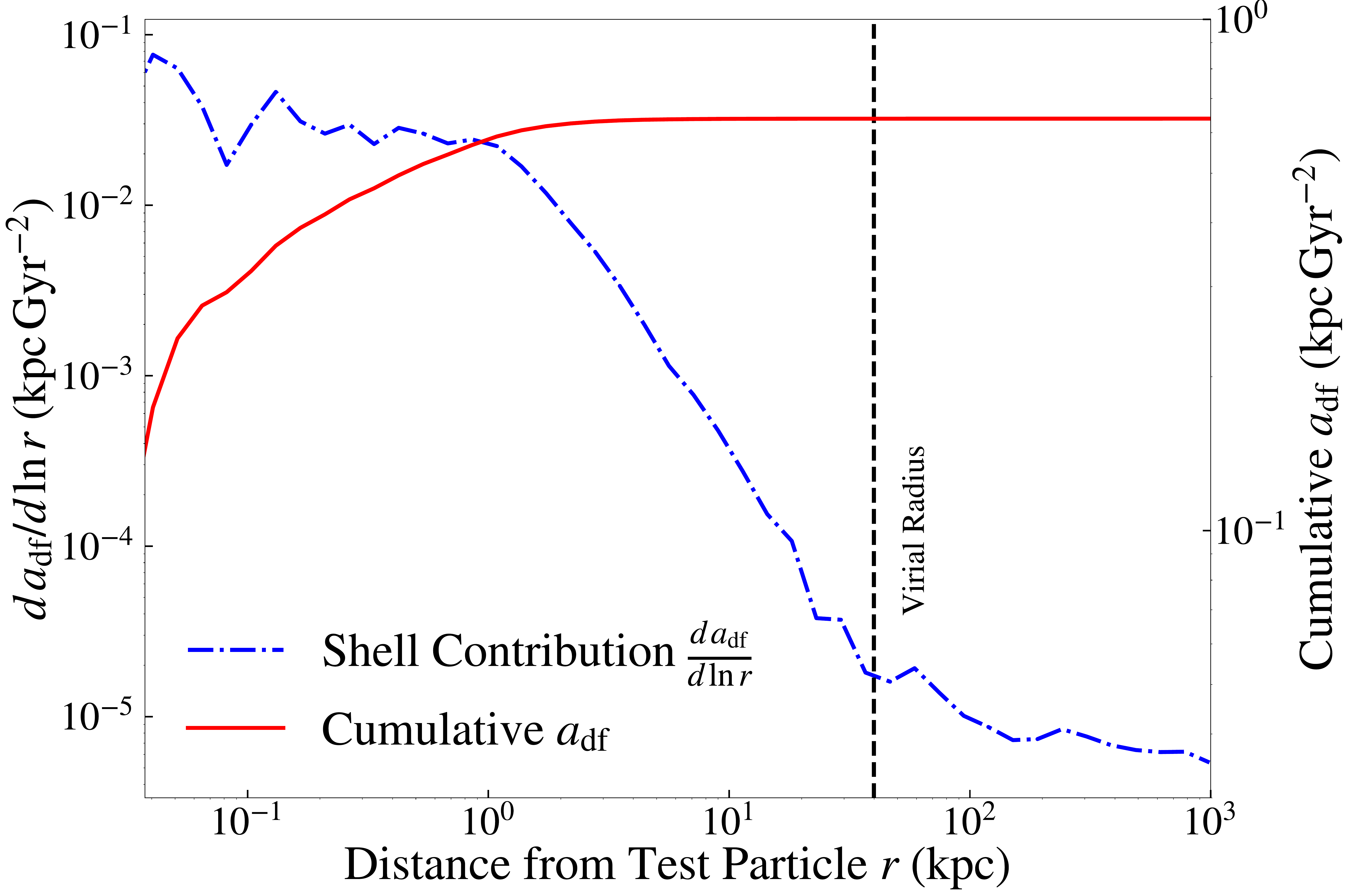}
    \caption{The contributions and cumulative results on DF from slices with different radius around a test particle. The test particle is a $10^5 M_\odot$ particle at $(0,1,0)$ with a $200\,\text{km/s}$ velocity in the $y$ direction. The particle's distance to the virial radius where we cutoff the sum is labelled with a black dashed line. We see that contributions are mostly from slices near the particle, while those from slices outside the virial radius are $\gtrsim1000$ times lower, suggesting that our cutoff makes little difference.}
    \label{fig:lss_slice}
\end{figure}

In Fig.~\ref{fig:df_compare_amp} we compare the acceleration amplitude $a_\mathrm{df} \equiv |{\bf a}_{\rm df}|$ calculated from different formulae for a test particle of mass $10^5\,M_\odot$ in the representative snapshot. The test particle is placed along an arbitrary $x$-axis passing through the galaxy center with a simulation-frame velocity of ${\bf V}_{M}=200\,{\rm km\,s^{-1}}\,\hat{y}$ (Fig. \ref{fig:galaxy}, red dashed line and arrow). We compare the results from our ``full'' expression (Eq.~\ref{eqn:discrete_df_smoothing}), our expression ignoring force softening (Eq.~\ref{eqn:discrete_df}), and the classical C43 expression (Eq. \ref{eqn:chandra_original}). Eqs.~\ref{eqn:discrete_df_smoothing} \&\ \ref{eqn:discrete_df} can be directly applied to the simulations without any processing. To apply Eq.~\ref{eqn:chandra_original}, we estimate the continuous $\rho$ at each position ${\bf x}_{M}$ using a kernel density estimator by averaging through the $0.4\,\mathrm{kpc}$ cubic box around ${\bf x}_{M}$; we calculate the local velocity integral by converting it into the usual discrete sum in this box, and we take $\ln{\Lambda}=5$ to be constant, once again fitting it so that the median/mean acceleration is essentially identical.

The agreement between Eq.~\ref{eqn:discrete_df_smoothing} and Eq.~\ref{eqn:chandra_original} is reasonable, but again this requires choosing $\Lambda$ specific for the problem and snapshot (we note, for example, that the effective $\Lambda$ here differs by almost a factor of two from the value fit to the idealized \citet{hernquist1990} profile sphere tests in the previous section). Eq.~\ref{eqn:discrete_df}, which ignores force softening, is also quite similar, except for occasional ``spikes'' arising from close proximity to $N$-body particles producing a spurious large force which is not actually present in the simulations (accounted for correctly in our Eq.~\ref{eqn:discrete_df_smoothing}).

Fig.~\ref{fig:df_compare_dir} similarly compares the direction $\hat{\bf a}_{\rm df}$. Because C43 assume homogeneity, and their ${\bf a}_{\rm df}^{\rm C43}$ has equal contributions from all scales, a major ambiguity in Eq.~\ref{eqn:chandra_original} -- even after we fit out the Coulomb logarithm -- is where/how to evaluate ${\bf V}={\bf V}_M-{\bf V}_m$. Should we interpolate to the local value at ${\bf x}_{M}$, weight by contribution to $\Lambda$, or weight by mass (dominated by distant particles)? If we follow the same procedure above to obtain a ``local'' ${\bf V}$, then we see that usually, the direction of $\hat{\bf a}_{\rm df}$ from Eq.~\ref{eqn:discrete_df_smoothing} and from Eq.~\ref{eqn:chandra_original} agree, especially if we assume a test particle $M$ with large lab-frame $|{\bf V}_{M}|$ (since then ${\bf V} \approx {\bf V}_{M}$, independent of the background ${\bf v}_{m}$). But when ${\bf V}_{M}$ is small (the case of interest for sinking), Eq.~\ref{eqn:chandra_original} can occasionally ``flip'' to point in an unphysical direction in a noisy velocity field.

Our Eq.~\ref{eqn:discrete_df_smoothing} allows us to easily quantify the contributions to the total $a_{\rm df}$ from all the mass in radial shells. Fig.~\ref{fig:lss_slice} shows this (specifically $d a_{\rm df} / d\ln{r}$, integrating the contributions from all particles in logarithmically-spaced shells of distance $r$ from $M$) again for a representative example (with $M$ at $|{\bf x}_{M}| = 1\,$kpc from the origin on the $x$-axis) in the same snapshot. At small scales ($r\lesssim 1\,$kpc) around $M$, where the density field is {\em statistically} homogeneous (there are local fluctuations, but there is not a strong systematic dependence of density on distance $r$ from $M$), we see the expected Coulomb log behavior ($d a_{\rm df} / d\ln{r} \sim $\,constant). At larger radii, the contribution falls rapidly. We can, for example, truncate the sum in Eq.~\ref{eqn:discrete_df_smoothing} at the virial radius (labeled) with negligible loss of accuracy. This is expected if the galaxy follows a realistic density profile, as in e.g.\ an isothermal sphere, the density is not constant, but at $r \gg |{\bf x}_{M}|$ falls rapidly ($\propto r^{-2}$, giving rapid convergence). As expected, the behavior at larger $r$ does motivate the value of $\ln{\Lambda}$ we fit: if we take $\Lambda=b_{\rm max}/b_{\rm min}$, with $b_{\rm min} \sim {\rm max}[(m/\rho)^{1/3},\,G\,M/V^{2}] \sim 1\,$pc, and $b_{\rm max} \sim 1\,$kpc, we obtain $\ln{\Lambda} \sim 7$, similar to our fitted value.

The above discussions are closely related to cases where the background field particles have a non-negligible physical bulk motion, like a wandering BH in a rotating disc-galaxy setup. While studying such simulations in detail is beyond the scope of this work, we comment that the rotating of star particles in the disc could largely affect the strength and direction of DF, since their phase space distribution departs significantly from homogeneity and isotropy. In an extremely dense galactic environment, we may expect the local disc particles with similar circular velocities contribute most to the BH's DF, such that the BH is boosted by the field particles around it, which is similar to the case we already studied. For a less dense setup, non-local (halo) particles with different circular velocity could be important, and their combined contribution to DF with local disc particles could make the BH dynamics more complicated. Our DF estimator, which applies to an arbitrary phase space distribution and counts the DF contribution from each individual field particle, would be ideal for studying such problems. Such topics will be studied in future work.

Briefly, one might wonder whether on sufficiently large scales, where the Universe becomes homogeneous and isotropic, $a_{\rm df}$ might begin to grow logarithmically again. However, even if we ignore finite speed-of-gravity effects (i.e.\ consider pure Newtonian gravity), on these scales the velocity must include the Hubble flow, so ${\bf v}_{\rm physical} = {\bf v}_{\rm peculiar} + H(z)\,{\bf r}$. In an isotropic pure Hubble-flow medium, the DF is identically zero, as there is always equal-and-opposite contributions to ${\bf a}_{\rm df}$ from the fact that ${\bf V} \propto {\bf r}$ (i.e.\ because $\langle {\bf V}\rangle=0$ on all scales). If we consider a Hubble flow plus peculiar velocities, then expanding Eq.~\ref{eqn:discrete_df_smoothing} appropriate for large $r$ where $\langle \rho(r) \rangle \sim $\,constant and $H\,r \gg \langle |{\bf v}_{\rm peculiar}(r)|^{2} \rangle^{1/2}$, the contributions to the sum take the form $\sum\,G^{2}\,M\,\langle |{\bf v}_{\rm peculiar}(r)|^{2} \rangle^{1/2}\,\Delta m_{i} / H^{3}\,r^{6} \propto \int \rho\,r^{-6}\,d^{3}{\bf x}$, which converges rapidly as $r\rightarrow \infty$.

\subsection{Interpolating the Sub-Grid Model in Simulations With Variable Masses}
\label{sec:double-count}

Finally, one can easily imagine situations such as cosmological simulations with a range of BH masses where the DF forces are well-resolved for some targets (e.g.\ supermassive BHs with $M_{\rm BH}\sim 10^{10}\,M_{\odot}$) but not others (e.g.\ lower-mass BHs). In these cases applying Eq.~\ref{eqn:discrete_df_smoothing} to all BHs would ``double count'' for some. A simple (albeit ad-hoc) approach to avoid double-counting is to multiply $\Delta {\bf a}_{\rm df}^{i}$ by a sigmoid or ``switch''-like function $g(\Delta m_{i}/M_{j},...)$ which has the property $g(x,...)\rightarrow 0$ for $x\rightarrow 0$ and $g(x,...) \rightarrow 1$ for $x\rightarrow \infty$. It is beyond the scope of our paper here to develop and test such models, and from Fig.~\ref{fig:values_evo} we see one complication is that this should depend on how one treats the force softening (not just particle masses $\Delta m_{i}$), but a quick examination of the idealized tests in \S~\ref{sec:results_sim} with different $M_{\rm BH}$ suggests (if we assume $\epsilon \sim \Delta x$, as usually adopted in such simulations) a simple function like $g =  \mathrm{min} (1.0, \mathrm{max}(0, (3/\log(M_{{\rm BH},\,j}/\Delta m_i)-1)/1.6))$ works reasonably well. Another advantage of our Eq.~\ref{eqn:discrete_df_smoothing} is that because it operates in pairwise fashion, it can naturally deal with simulations with a wide range of $\Delta m_{i}$ (a common situation), while attempting to apply such a correction factor ``locally'' to Eq.~\ref{eqn:chandra_original} leaves it ill-defined which value of $\Delta m_{i}$ to use.

\section{Conclusions} \label{sec:discussion}

In numerical simulations, especially of star and galaxy formation, it is common to encounter the limit where DF {\em should} be experienced ($M\gtrsim m$) by some explicitly-evolved objects $M$ (e.g.\ black holes, massive stars), but it cannot be {\em numerically} resolved ($\Delta m_{i} \gtrsim M$). As a result, there have been several attempts to develop and apply ``on the fly'' sub-grid DF models. Almost all of these amount to some attempt to calculate and apply the traditional C43 formula (Eq.~\ref{eqn:chandra_original}) to the masses $M$ at each time \citep[see e.g.][]{colpi:2007.binary.in.mgrs,dotti:bh.binary.inspiral,Tremmel2015,Pfister2019}. However, this can introduce a number of problems in practice, namely the ambiguity of kernel-dependent locally-defined quantities, inconsistency in applying force softening and momentum conservation, the semi-arbitrary choice of Coulomb logarithm, the necessity of assuming Maxwellian velocity distribution functions, and additional computational expenses for kernel estimates.

In this manuscript, we derive a new discrete expression for the DF force, ${\bf a}_{\rm df}$, given in Eq.~\ref{eqn:discrete_df_smoothing}. This formula is specifically designed for application to numerical simulations, either in post-processing, or ``on the fly'' when the DF forces cannot be resolved (e.g.\ when $N$-body particle masses are comparable to the BH mass $M$, as a ``sub-grid'' DF model). While still approximate, this has a large number of advantages compared to the traditional \citet{chandrasekhar1943} (C43) analytic expression, including (1) it allows for an arbitrary distribution function, without requiring an infinite homogeneous time-invariant medium with constant density, Maxwellian velocity distribution, etc. (but it does reduce {\em identically} to a discrete form of the C43 expression, when these assumptions are actually satisfied); (2) it is designed specifically for simulations so it is represented only as a sum over quantities which are always well-defined in the simulation for all $N$-body particles (e.g.\ positions, velocities, masses), and does not require the expensive and fundamentally ill-defined evaluation of quantities like a ``smoothed'' density, background mean velocity/dispersion/distribution function, Coulomb logarithm, etc.; (3) it trivially incorporates force softening exactly consistent with how it is treated in-code, and generalizes to arbitrary multi-component $N$-body simulations with different species and an arbitrary range of particle masses; (4) it manifestly conserves total momentum, unlike $N$-body implementations of C43; (5) it can be evaluated directly alongside the normal gravitational forces with negligible cost, and automatically inherits all of the desired convergence and accuracy properties of the $N$-body solver. We have implemented this ``live'' evaluation of Eq.~\ref{eqn:discrete_df_smoothing} in {\small GIZMO}, and verified that all of the properties above apply, that it agrees well with our $N$-body simulations, and that the computational overhead of evaluating it alongside gravity in the tree is immeasurably small.

There are still uncertainties in our work. In our derivation of the discrete formula, we inserted an approximate integral kernel, which is not necessarily unique or best-behaved. We found that even if our discrete estimator closely agrees with the calibrated-Chandrasekhar DF estimator in our test problems, it still differs from the the high-resolution simulation results in terms of the detailed particle trajectories, which might be related to the fundamental Chandrasekhar-like assumptions we have made in our formula. We also note that it remains an open question how to accurately avoid ``double counting'' when some of the DF may be captured self-consistently by the $N$-body code while additional DF is modeled using our sub-grid model. This is especially the case when the system evolves (such as when supermassive black holes grow) and the fraction of ``resolved'' dynamical friction changes with time. Future work will be needed to make improvements on these points.

\section{Acknowledgements}

We thank Sophia Taylor for early contributions to the development of the discrete dynamical friction estimator, and Daniel Anglés-Alcázar for useful discussions. Support for LM \&\ PFH was provided by NSF Collaborative Research Grants 1715847 \&\ 1911233, NSF CAREER grant 1455342, NASA grants 80NSSC18K0562, JPL 1589742. LZK was supported by NSF-AAG-1910209, and by the Research Corporation for Science Advancement through a Cottrell Fellowship Award.  CAFG was supported by NSF through grants AST-1715216, AST-2108230, and CAREER award AST-1652522; by NASA through grants 17-ATP17-0067 and 21-ATP21-0036; by STScI through grants HST-AR-16124.001-A and HST-GO-16730.016-A; by CXO through grant TM2-23005X; and by the Research Corporation for Science Advancement through a Cottrell Scholar Award. Numerical calculations were run on the Caltech compute cluster ``Wheeler,'' allocations FTA-Hopkins supported by the NSF and TACC, and NASA HEC SMD-16-7592.

\section*{Data Availability}
The data and source code supporting the plots within this article are available on reasonable request to the corresponding author.

\bibliographystyle{mnras}
\bibliography{df.bib}
%%%%%%%%%%%%%%%%%%%%%%%%%%%%%%%%%%%%%%%%%%%%%%%%%%

% Don't change these lines
\bsp	% typesetting comment
\label{lastpage}
\end{document}